\documentstyle[prl,aps,multicol,epsf]{revtex}
\tighten
\begin{document}
\draft
\title{The $g$-factors of discrete levels in nanoparticles}
\author{K.A. Matveev$^{1}$, L.I. Glazman$^{2}$,  and A.I. Larkin$^{2,3}$}
\address{
$^{1}$Department of Physics, Duke University, Durham, NC 27708-0305\\
$^{2}$Theoretical Physics Institute, University of Minnesota, 
Minneapolis, MN 55455\\
$^{3}$L.D. Landau Institute for Theoretical Physics, 117940 Moscow, 
Russia\\}
\maketitle

\begin{abstract}
  Spin-orbit scattering suppresses Zeeman splitting of individual energy
  levels in small metal particles.  This suppression becomes significant
  when the spin-orbit scattering rate $\tau_{\rm so}^{-1}$ is comparable
  with the quantum level spacing $\delta$.  The $g$-factor exhibits
  mesoscopic fluctuations; at small $\delta\tau_{\rm so}$ it is
  distributed according to the Maxwell distribution. At $\delta\tau_{\rm
    so}\to 0$ the average $g$-factor levels off at a small value $g\sim
  (l/L)^{1/2}$ given by the ratio of the electron mean free path $l$ to
  the particle size $L$.  On the contrary, in 2D quantum dots the
  $g$-factor is strongly enhanced by spin-orbit coupling.
\end{abstract}

\begin{multicols}{2}

In bulk metals the spin-orbit interaction is known to affect the
$g$-factor of electrons.  These corrections are relatively
weak\cite{Halperin}; even in such a heavy metal as gold, the $g$-factor is
2.1, and differs only slightly from the nominal value of 2.  On the
contrary, in recent experiments\cite{Black} with Al nanoparticles the
measured values of the $g$-factor were in the interval between 0.27 and
1.9.  Later a similar effect was observed in gold
nanoparticles\cite{Davidovic}, where the value $g\approx 0.3$ was
measured.  A systematic study of the effect of spin-orbit scattering on
the $g$-factors of discrete levels was undertaken in
Ref.~\onlinecite{Ralph}.  In particular, it was demonstrated that when 4\%
of gold impurities are added to Al particle, the $g$-factor drops from 2
to about~0.7.

In bulk materials the $g$-factor is conventionally determined from the
electron spin resonance (ESR) data. ESR involves transitions between
states of the electron continuum. On the contrary, the experiments with
nanoparticles\cite{Black,Davidovic,Ralph} study the Zeeman splitting of
individual electron levels $\epsilon_i$:
\begin{equation}
  \label{eq:Zeeman}
  \epsilon_{i\sigma}(H) = \epsilon_i \pm \frac12  g_i \mu_B H.
\end{equation}
Here $H$ is the magnetic field, $\mu_B=e\hbar/2mc$ is Bohr magneton; $m$
is the free electron mass.  The splitting (\ref{eq:Zeeman}) is linear in
$H$ as long as it is small compared to the quantum level spacing $\delta$.
The $g$-factor determined by Eq.~(\ref{eq:Zeeman}), in general, varies
from level to level.  In the absence of spin-orbit interaction $g_i=2$.
Evidently\cite{Black,Davidovic,Ralph} the spin-orbit scattering affects
the values of $g_i$.

In this paper we develop a theory of the $g$-factors of individual levels.
Unlike its bulk value, the $g$-factor defined by Eq.~(\ref{eq:Zeeman}) is
very sensitive to even weak spin-orbit interaction.  The reason is that
the magnitude of the correction to $g=2$ value measured by ESR in the bulk, is
determined by the comparison of spin-orbit interaction to the Fermi energy
$E_F$, whereas the relevant energy scale for $g$-factors of individual
levels is the quantum level spacing $\delta\ll E_F$.  Indeed, the
spin-orbit interaction is usually described by the mean time of spin-orbit
scattering $\tau_{\rm so}$.  The scattering time should be compared with
the time $\sim1/\delta$ that electron travels along a closed trajectory
corresponding to a quantum level.  At $\delta\tau_{\rm so}\gg1$ the
electron spin flips very infrequently, and the effect of the spin-orbit
scattering is weak.  On the contrary, at $\delta\tau_{\rm so}\ll1$ the
spin flips the average of $N=1/\delta\tau_{\rm so}$ times during the
electron motion along the closed trajectory.  Thus the average spin in
such a quantum state is significantly less than 1/2, and the response
(\ref{eq:Zeeman}) to the magnetic field is strongly suppressed.  The
$g$-factor of this level can be estimated by assuming that electron
spin-flips occur at random moments in time.  The root-mean-square value of
the spin is then $1/\sqrt{N}$, resulting in $g_i \sim
\sqrt{\delta\tau_{\rm so}}$.

The above estimate accounts only for the spin contribution to the
$g$-factor.  In fact, the level splitting in a magnetic field is
determined by the total magnetic moment of the state
\begin{equation}
  \label{eq:momenta}
  \langle M\rangle_i = \mu_B^* \langle l_z \rangle_i 
                       + 2 \mu_B \langle s_z \rangle_i.
\end{equation}
Here $l_z$ and $s_z$ are the operators of the angular momentum and spin,
respectively; $\mu_B^*=e\hbar/2m^*c$, with $m^*$ being the effective mass
of the electrons.  The brackets $\langle\cdots\rangle_i$ denote the
expectation value in $i$-th eigenstate of an electron confined to a
nanoparticle in an infinitesimal magnetic field in $z$-direction. In a
nanoparticle of a generic shape, the orbital levels are not degenerate.
Then in the absence of the spin-orbit interaction the time reversal
symmetry dictates $\langle l_z \rangle_i = 0$ at $H\to0$.  It is important
to note that in the presence of spin-orbit interaction $\langle l_z
\rangle_i \neq 0$, and the orbital motion contributes to the splitting of
levels by magnetic field\cite{Kravtsov}.

One can estimate the angular momentum of the electron as the (directed)
area $A$ covered by its trajectory divided by the period of motion
$1/\delta$, that is, $\langle l_z \rangle_i\sim m^* A\delta$.  To find $A$
we notice that during time $1/E_T$ the electron travels across the grain
and, therefore, covers the area $\sim L^2$.  (Here $E_T=D/L^2$ is the
Thouless energy, $L$ is the grain size, and $D$ is the diffusion constant
for electrons in the grain.)  During the period of motion $1/\delta$ the
electron bounces off the boundaries $\sim E_T/\delta$ times.  Since
the direction of motion after each bounce is random, the total directed
area is $A\sim L^2 \sqrt{E_T/\delta}$.  Consequently, the root-mean-square
angular momentum of each state is
\begin{equation}
  \label{eq:l_z}
  \langle l_z \rangle_i\sim m^* L^2 \sqrt{E_T\delta}
    \sim\begin{cases}{
         \sqrt{l/L}, &   \mbox{3D},\cr
         \sqrt{k_F l},&  \mbox{2D}.}
        \end{cases}
\end{equation}
Here $l$ is the transport mean free path of electrons, and $k_F$ is their
Fermi wavevector.  

One can also see from Eq.~(\ref{eq:l_z}) that in a three-dimensional (3D)
nanoparticle the orbital contribution to the $g$-factor is small $g_i^{\rm
  orb} \sim \sqrt{l/L}\ll2$.  However, in the case of strong spin-orbit
scattering, $\delta\tau_{\rm so}\ll1$, the above estimated spin
contribution is also small, $g_i^{\rm sp}\sim \sqrt{\delta\tau_{\rm so}}
\ll2$.  Therefore in this regime both contributions may have to be taken
into account.

We will now show that at $\delta\tau_{\rm so}\ll1$ the $g$-factor of a
level is a random quantity with the distribution function
\begin{equation}
  \label{eq:distribution}
  P(g) = 3\sqrt{\frac{6}{\pi}}
         \frac{g^2}{\langle\!\langle g^2\rangle\!\rangle^{3/2}}
         \exp\left(-\frac{3g^2}
                    {2\langle\!\langle g^2\rangle\!\rangle}\right),
\end{equation}
where the averaging $\langle\!\langle \cdots\rangle\!\rangle$ is performed
either over an ensemble of 3D nanoparticles or over different levels in a
single nanoparticle. Furthermore, we express $\langle\!\langle
g^2\rangle\!\rangle$ in terms of two quantities, $\tau_{\rm so}$ and $l$,
which can be measured independently:
\begin{equation}
  \label{eq:gsquared}
  \langle\!\langle g^2\rangle\!\rangle =
  \frac{6}{\pi\hbar}\delta\tau_{\rm so} +  \alpha \frac{l}{L}.
\end{equation}
Here the dimensionless constant $\alpha$ is determined by the geometry of
the nanoparticle; its exact value will be discussed later.  Eqs.
(\ref{eq:distribution}) and (\ref{eq:gsquared}) are the main result of
this paper.

We begin the study of the $g$-factors of individual levels by finding
relation between $g_i$ and the matrix elements of the operator of magnetic
moment $M$.  Due to the time-reversal symmetry, the levels in the
nanoparticle are degenerate, with the wavefunctions $|\psi_i\rangle$ and
$|T\psi_i\rangle$, where $T$ stands for the time reversal operator.  In a
small magnetic field $H$ the levels are split by the perturbation $MH$.
Using the standard method of degenerate perturbation theory, one can
easily find the splitting in the form (\ref{eq:Zeeman}) with the
$g$-factor
\begin{equation}
  \label{eq:secular}
  g_i=2\frac{|\vec\mu|}{\mu_B},
\end{equation}
where the real vector $\vec\mu$ is defined as
\begin{equation}
  \label{eq:mu}
  \mu_x+i\mu_y=\langle T\psi_i|M|\psi_i\rangle,
  \quad \mu_z=\langle \psi_i|M|\psi_i\rangle.
\end{equation}
The distribution function $p(\vec\mu)$ is, by definition, 
\begin{eqnarray}
  p(\vec\mu)=\int\frac{d^3\lambda}{(2\pi)^3} e^{i\vec\lambda\cdot\vec\mu} 
      \langle\!\langle
      \exp(-i\lambda_x{\rm Re}\langle T\psi_i|M|\psi_i\rangle
\nonumber\\
           -i\lambda_y{\rm Im}\langle T\psi_i|M|\psi_i\rangle
           -i\lambda_z\langle \psi_i|M|\psi_i\rangle)
      \rangle\!\rangle.
  \label{eq:mu-distribution}
\end{eqnarray}

To perform the averaging in Eq.~(\ref{eq:mu-distribution}), we use the
Random Matrix Theory (RMT) approach \cite{Porter}.  Instead of the
ensemble of nanoparticles with strong spin-orbit scattering we will
consider an ensemble of symplectic matrices of size $2N\times 2N$ with
$N\gg1$.  Then the eigenfunctions $\langle\psi|$ and $\langle T\psi|$ of
the Hamiltonian are $N$-component spinors:
\begin{equation}
  \label{eq:spinors}
  \langle\psi| = (\{\phi_k^*,\chi_k^*\}),
  \quad
  \langle T\psi| = (\{-\chi_k,\phi_k\}).
\end{equation}

The operator of magnetic moment $M$ is a hermitian matrix, which due to
its time-reversal properties can be diagonalized to the form ${\rm
  diag}\{M_1,-M_1,\ldots, M_k,-M_k, \ldots, M_N,-M_N\}$.  In the basis of the
eigenfunctions of $M$ the matrix elements (\ref{eq:mu}) take the form:
\begin{eqnarray}
  \label{eq:matrix_elements}
  \langle\psi|M|\psi\rangle&=&\sum_{k=1}^N M_k(|\phi_k|^2-|\chi_k|^2),
  \\
  \langle T\psi|M|\psi\rangle&=&-2\sum_{k=1}^N M_k\phi_k\chi_k.
\end{eqnarray}
The advantage of the RMT approach is that the ensemble averaging in
Eq.~(\ref{eq:mu-distribution}) is easily performed using the Porter-Thomas
distribution\cite{Porter} of the matrix elements:
\begin{equation}
  \label{eq:porter-thomas}
  \langle\!\langle\cdots\rangle\!\rangle =
  \int\prod_{k=1}^N \frac{d^2\phi_k d^2\chi_k}{(\pi/2N)^2}
    e^{-2N(|\phi_k|^2+|\chi_k|^2)}\cdots.
\end{equation}
The averaging in Eq~(\ref{eq:mu-distribution}) with the help of
(\ref{eq:matrix_elements})--(\ref{eq:porter-thomas}) reduces to the
calculation of $N$ identical quadruple Gaussian integrals.  The result has
the form 
\begin{equation}
  \label{eq:distribution-mu}
    p(\vec\mu)=\int\frac{d^3\lambda}{(2\pi)^3}
    e^{i\vec\lambda\cdot\vec\mu}
    \prod_{k=1}^N \left(1+\frac{|\vec\lambda|^2M_k^2}{(2N)^2}\right)^{-1}.
\end{equation}
In the limit of large $N\gg1$, this integral becomes Gaussian too, and we
find
\begin{equation}
  \label{eq:pmu}
  p(\vec\mu)=\left(\frac{2N^2}{\pi{\rm Tr}M^2}\right)^{3/2}
              \exp\left(-\frac{2N^2}{{\rm Tr}M^2}|\vec\mu|^2\right).
\end{equation}
Taking into account the relation (\ref{eq:secular}), we now immediately
find the distribution function of the $g$-factor in the form
(\ref{eq:distribution}) with the mean square $g$ defined as
\begin{equation}
  \label{eq:gsquared_RMT}
  \langle\!\langle g^2\rangle\!\rangle = 
      \frac{3}{\mu_B^2}  \frac{{\rm Tr} M^2}{N^2}.
\end{equation}

As a phenomenological theory, RMT enabled us to find the functional form
of the distribution function; however, the width of the distribution
(\ref{eq:gsquared_RMT}) is now expressed in terms of a phenomenological
parameter ${\rm Tr} M^2/N^2$.  The relation between this parameter and the
microscopic properties of the system cannot be established within the RMT.
To do this, one has to find an observable quantity $Q$ which can be
evaluated within both the phenomenological RMT and a microscopic theory.

We choose $Q$ to be the energy absorbed in unit time from a weak external
ac magnetic field $H(t)=H_0\cos\omega t$.  Such a field induces
transitions between quantum levels in the grain, leading to the absorption
of the energy in the grain.  Within the RMT, the absorption can be found
with the help of the Fermi golden rule as
\begin{equation}
  \label{eq:golden}
  Q=\left\langle\!\!\!\left\langle2\pi\!\!\sum_{k\leq k_f<p}
      \left|{\textstyle\frac12}H_0\langle\psi_k|M|\psi_p \rangle\right|^2
       \delta(\omega+\epsilon_k-\epsilon_p)
     \right\rangle\!\!\!\right\rangle\omega.
\end{equation}
The expression inside $\langle\!\langle\ldots\rangle\!\rangle$ is the rate
of absorption of quanta of radiation interacting with the system.  The
absorption occurs due to the transitions from occupied states with $k\leq
k_f$ to the empty states, $p>k_f$, and are induced by the term $\frac12
H_0 M e^{-i\omega t}$ in the corresponding coupling Hamiltonian.

In RMT the energy levels and the eigenfunctions are uncorrelated, i.e.,
the averaging over the energy levels and matrix elements in
Eq.~(\ref{eq:golden}) can be performed independently.  Also, at
$\omega\gg\delta$, one can neglect the correlations of the densities of
states at energies separated by $\omega$, and we find
\begin{equation}
  \label{eq:golden_2}
  Q=\frac{\pi\omega^2H_0^2}{2\delta^2}
    \langle\!\langle |M_{kp}|^2\rangle\!\rangle.
\end{equation}
It is important to note that contrary to the above calculation of the
$g$-factors of individual levels, Eqs.~(\ref{eq:secular}) and
(\ref{eq:mu}), here the matrix elements $M_{kp}$ are between eigenstates
with different energies.  Upon averaging over the ensemble
$\langle\!\langle |M_{kp}|^2\rangle\!\rangle$ becomes independent of $k$
and $p$.  Its magnitude can be found by presenting the invariant ${\rm Tr}
M^2$ as a sum of $\langle\!\langle |M_{kp}|^2\rangle\!\rangle$ over all
$2N$ values of $k$ and $p$.  Since the number of diagonal matrix elements
$2N$ is small compared to the number of the off-diagonal ones, $4N^2-2N$,
we conclude that $\langle\!\langle |M_{kp}|^2\rangle\!\rangle={\rm Tr}
M^2/4N^2$ at $N\gg1$.  Combining this relation with
Eqs.~(\ref{eq:golden_2}) and (\ref{eq:gsquared_RMT}), we find
\begin{equation}
  \label{eq:gsquared_Q}
  \langle\!\langle g^2\rangle\!\rangle = 
    \frac{24\,\delta^2}{\pi\hbar\omega^2(\mu_BH_0)^2}\,Q.
\end{equation}
It is noteworthy that this result, obtained within the RMT, contains no
phenomenological parameters. It establishes a relation between the
property of a single level, the $g$-factor, and a macroscopic quantity $Q$
insensitive to the effects of discreteness of levels.

To evaluate the absorption rate $Q$ for a given nanoparticle, it is
convenient to express it as $Q=\frac12\omega A'' H_0^2$ in terms of the
imaginary part $A''$ of the $zz$-component of the tensor of magnetic
polarizability of the sample $A_{ik}$, defined as $M_i=A_{ik}H_k$,
Ref.~\onlinecite{Landau}.  Then the result for the mean square $g$-factor
takes the form
\begin{equation}
  \label{eq:polarizability}
  \langle\!\langle g^2\rangle\!\rangle = 
    \frac{12\,\delta^2}{\pi\hbar\mu_B^2}\,\frac{A''(\omega)}{\omega}.
\end{equation}
It is well known\onlinecite{Landau} that at $\omega\to0$ the imaginary
part of the polarizability vanishes as $A''(\omega)\propto\omega$.  

Since $M=\mu_B^*l_z+2\mu_B s_z$, one can distinguish between the orbital and
spin contributions to the magnetic polarizability of the nanoparticle.
The spin contribution has the form
\begin{equation}
  \label{eq:spin_polarizability}
  A_s(\omega) = \frac{\mu_B^2\delta^{-1}}{1-i\omega\tau_{\rm so}/2}.
\end{equation}
Here the numerator is the usual static Pauli susceptibility of the
electron gas of the nanoparticle, and the denominator accounts for the
fact that  spin correlations decay exponentially with the time constant
$\tau_{\rm so}/2$. Substituting the imaginary part of the polarizability
(\ref{eq:spin_polarizability}) into (\ref{eq:polarizability}), we
reproduce the first term of our main result (\ref{eq:gsquared}).

The orbital contribution to the magnetic polarizability is due to the
magnetic moment of the eddy currents generated in the sample by the ac
magnetic field.  For a particle of a shape symmetrical with respect to the
rotations around the $z$-axis one can easily find
\begin{equation}
  \label{eq:eddy}
  A''= \frac{\omega\sigma}{4c^2}\, {\overline {\rho_\perp^2}}\, V,
  \quad
  \omega\to0.
\end{equation}
Here $\sigma$ is the conductivity of the metal, $V$ is the volume of the
nanoparticle, and ${\overline {\rho_\perp^2}}$ is the ``moment of
inertia'' of the grain, assuming unit density.  For a spherical
nanoparticle of radius $L$ the combination of
Eqs.~(\ref{eq:polarizability}) and (\ref{eq:eddy}) reproduces the second
term in Eq.~(\ref{eq:gsquared}) with $\alpha=(6/5)(m/m^*)^2$.

For the thin ring geometry Eqs.~(\ref{eq:polarizability}) and
(\ref{eq:eddy}) result in $\langle\!\langle g^2\rangle\!\rangle=
3m^2L^2D\delta/\pi^3\hbar^3$, where $L$ stands for the circumference of
the ring.  In the context of the persistent current problem the orbital
effect of magnetic field on the splitting of the energy levels was studied
earlier by Kravtsov and Zirnbauer\cite{Kravtsov}.  They used the
non-linear $\sigma$-model techniques \cite{Efetov} to solve the general
problem of crossover from the symplectic ensemble to the orthogonal one.
A special limiting case of that solution gave rise to a distribution
function of level splittings, which reassuringly coincides with our result
for the thin ring geometry.  Phenomenologically the crossover problem was
solved within RMT by Mehta and Pandey\cite{Pandey}.  Our approach enables
one to express their phenomenological crossover parameter to physical
observables without resorting to $\sigma$-model calculations.

In the case of weak spin-orbit interaction, $\delta\tau_{\rm so}\gg1$, the
correction to the average $g$-factor is small and can be found by
perturbation theory.  In the lowest order perturbation theory in
spin-orbit coupling one finds\cite{Halperin}
\begin{equation}
  \label{eq:perturbation}
  g_i=2-\frac{2\hbar\delta}{\pi\tau_{\rm so}}
       \sum_{j\neq i}\frac{1}{(\epsilon_i-\epsilon_j)^2}.
\end{equation}
\begin{figure}
\epsfxsize=0.41\textwidth
\hspace{1pt}
\epsffile{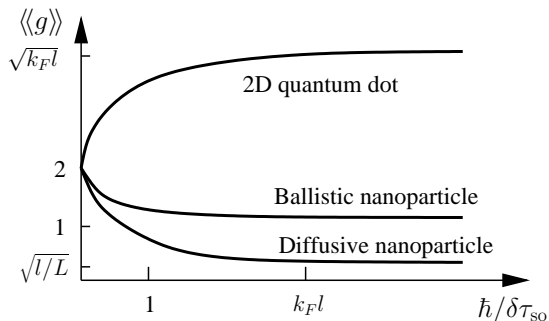} 
  \caption{Sketch of the dependence of the average $g$-factor on the
    strength of spin-orbit scattering, see text. We
    assumed $m^*/m\simeq 1$ in this plot.}    
    \label{fig:1}
\end{figure}

\noindent 
Assuming that the energy levels $\epsilon_j$ are equidistant, the sum
(\ref{eq:perturbation}) was evaluated by Kawabata\cite{Kawabata}.
However, in a disordered system the Wigner-Dyson statistics of the energy
levels is a more realistic assumption.  In this case one of the levels $j$
can be close to level $i$, resulting in a particularly large correction to
the $g$-factor.  Taking into consideration the fact that the level
repulsion in the orthogonal ensemble suppresses the probability $p_o$ of
two levels being very close, $p_o(\epsilon_i-\epsilon_j) =
\pi^2|\epsilon_i - \epsilon_j|/6 \delta^2$,\cite{Porter}, we find that the
average value of the sum in Eq.~(\ref{eq:perturbation}) is logarithmically
large:
\begin{equation}
  \label{eq:g-factor_perturbative}
  \langle\!\langle g \rangle\!\rangle
    =2-\frac{\pi}{6}\frac{\hbar}{\delta\tau_{\rm so}}
     \ln \frac{\delta\tau_{\rm so}}{\hbar}.
\end{equation}
The logarithmic divergence in Eq.~(\ref{eq:g-factor_perturbative}) was
cut-off at the energy scale
$|\epsilon_i-\epsilon_{j}|\sim\sqrt{\delta/\tau_{\rm so}}$, because of the
additional level repulsion caused by the weak spin-orbit coupling.

Our results are summarized in Fig.~\ref{fig:1}.  The lower curve shows the
dependence of the average $g$-factor on the strength of the spin-orbit
scattering in the case of a diffusive nanoparticle.  In the regime of
$1/\delta\tau_{\rm so}\ll1$ the behavior is described by
Eq.~(\ref{eq:g-factor_perturbative}), and at $1/\delta\tau_{\rm so}\gg1$
the average $g$-factor drops in agreement with Eq.~(\ref{eq:gsquared}).
In the latter regime $\langle\!\langle g \rangle\!\rangle =
(8/3\pi)^{1/2}\langle\!\langle g^2 \rangle\!\rangle^{1/2}$, as one can
easily see from the distribution function (\ref{eq:distribution}).  

The values of the mean free path are usually not well known in the
experiment, and it is possible that $l$ exceeds the grain size $L$.  The
middle curve in Fig.~\ref{fig:1} shows schematically the behavior of a
ballistic nanoparticle, where all the scattering of electrons is due to
the reflection from the boundaries.  Qualitatively, in this case one can
use the results for the diffusive case with the transport mean free path
replaced by the size of the grain.  It is then obvious from
Eq.~(\ref{eq:gsquared}) that $\langle\!\langle g \rangle\!\rangle$ levels
off at a value of order unity in the limit of strong spin-orbit
scattering, as illustrated in the figure.  One can conduct a quantitative
study in a simple model of a spherical ballistic nanoparticle of radius
$L$ with totally diffusive scattering off the boundaries.  The magnetic
polarizability can be calculated with the help of the Kubo formula, which
relates it to the correlation function of angular momenta at different
times. This correlation function can be evaluated by considering the
classical electron trajectories inside the nanoparticle, in the spirit of
Ref.~\onlinecite{DeGennes}.  The exact result amounts to replacing
$l\to5L/8$ in Eq.~(\ref{eq:gsquared}) and keeping $\alpha$ the same as in
the case of a diffusive sphere.  Taking into consideration the rapid
decrease of the distribution function (\ref{eq:distribution}) of the
$g$-factor at $g^2\ll \langle\!\langle g^2 \rangle\!\rangle$, one should
conclude that the nanoparticles showing the values of $g^2$ well below the
ballistic value $\langle\!\langle g^2 \rangle\!\rangle=(3/4)(m/m^*)^2$ are
most likely in the diffusive regime.

The top curve in Fig.~\ref{fig:1} shows the behavior of the $g$-factor in
the case of a 2D quantum dot in a perpendicular magnetic field.  In
accordance with Eq.~(\ref{eq:l_z}), in the strong spin-orbit scattering
case, $1/\delta\tau_{\rm so}\gg1$, the orbital moment $\langle l_z\rangle$
reaches a very large value $\sim\sqrt{k_F l}$.  In experiments with
quantum dots in GaAs heterostructures the effect of the orbital field
should be further enhanced due to a small effective mass $m^*\approx 0.067
m$ of electrons, so one can expect to find $\langle\!\langle g
\rangle\!\rangle \sim (m/m^*) \sqrt{k_Fl}$.  On the other hand, at weaker
spin-orbit scattering, $1/\delta\tau_{\rm so}\lesssim 1$, the orbital
enhancement of the $g$-factor is reduced, as the orbital motion does not
affect the $g$-factor at all in the absence of spin-orbit coupling. The
orbital contribution at $1/\delta\tau_{\rm so}\ll 1$ can be found within
the first-order perturbation theory in the spin-orbit coupling, resulting
in $\langle\!\langle g \rangle\!\rangle\sim (m/m^*) (\hbar
k_Fl/\delta\tau_{\rm so})^{1/2}$.  Thus, the $g$-factor of a 2D quantum
dot is dominated by the effects of the orbital motion of electrons at
$1/\delta\tau_{\rm so}\gtrsim (k_Fl)^{-1}(m^*/m)^2$, whereas at
$1/\delta\tau_{\rm so}\ll (k_Fl)^{-1}(m^*/m)^2$ values close to $g=2$
should be observed.

This work was supported by NSF under Grants DMR-9974435, DMR-9731756, and
DMR-9812340. KM also acknowledges support by A.P. Sloan Foundation and the
kind hospitality of the TPI at the University of Minnesota.  The authors
acknowledge the hospitality of ICTP in Trieste, Italy and Centre for
Advanced Study in Oslo, Norway, where part of this work was performed.  We
are grateful to B.L. Altshuler, K.B. Efetov, S. Gu\'eron, V.E.  Kravtsov,
and D.C.  Ralph for useful discussions.

\end{multicols}

\end{document}